\RequirePackage{etoolbox}
\csdef{input@path}{{style/}{graphics/}}
\documentclass[MSNbibl,nameyear,seceqn,dvips]{arxstspdf}
\usepackage{graphicx}
\usepackage{url,breakurl}

\usepackage{flushend}
\usepackage{stfloats}

\volume{30}
\issue{1}
\pubyear{2015}
\firstpage{40}
\lastpage{58}
\doi{10.1214/14-STS430} 

\makeatletter
\newcommand{\nfrac}[2]{#1/#2}
\newcommand{\abfrac}[2]{(#1/#2)}

\newproclaim{Definition}{Definition}[section]

\newproclaim{definition}{Definition}
\newproclaim{result}{Result}
\newproclaim{Result}{Result}
\makeatother

\begin{document}
\begin{frontmatter}

\title{Extropy: Complementary Dual of Entropy}
\runtitle{Extropy:
Complementary Dual of Entropy}

\begin{aug}
\author[A]{\fnms{Frank}~\snm{Lad}\corref{}\ead[label=e1]{F.Lad@math.canterbury.ac.nz}},
\author[B]{\fnms{Giuseppe} \snm{Sanfilippo}\ead[label=e2]{giuseppe.sanfilippo@unipa.it}}
\and
\author[C]{\fnms{Gianna} \snm{Agr\`o}\ead[label=e3]{gianna.agro@unipa.it}}
\runauthor{F. Lad, G. Sanfilippo and G. Agr\`o}

\affiliation{University of Canterbury, University of Palermo and University of Palermo}

\address[A]{Frank Lad is  Research Associate,
Department of Mathematics and Statistics,
 University of Canterbury,
Christchurch, 8020, New Zealand
 \printead{e1}.}

\address[B]{Giuseppe Sanfilippo is Assistant Professor,
Dipartimento di Matematica e Informatica,
 University of Palermo,
Viale Archirafi 34, Palermo 90123, Italy
\printead{e2}.}

\address[C]{Gianna Agr\`{o} is  Associate Professor,
Dipartimento di Scienze Economiche, Aziendali e Statistiche,
University of Palermo,
Viale delle Scienze ed. 13,
Palermo 90128, Italy
\printead{e3}.}
\end{aug}

\begin{abstract}
This article provides a completion to theories of information based
on entropy, resolving a longstanding question in its axiomatization
as proposed by Shannon and pursued by Jaynes. We show that Shannon's
entropy function has a complementary dual function which we call
``extropy.'' The entropy and the extropy of a binary distribution
are identical. However, the measure bifurcates into a pair of
distinct measures for any quantity that is not merely an event
indicator.  As with entropy, the maximum extropy distribution is
also the uniform distribution,  and both measures are invariant with
respect to permutations of their mass functions. However, they
behave quite differently in their assessments of the refinement of a
distribution, the axiom which concerned Shannon and Jaynes. Their
duality is specified via the relationship among the entropies and
extropies of course and fine partitions. We also analyze the extropy
function for densities, showing that relative extropy constitutes a
dual to the Kullback--Leibler divergence, widely recognized as the
continuous entropy measure. These results are unified within the
general structure of Bregman divergences. In this context they
identify half the $L_2$ metric as the extropic dual to the entropic
directed distance. We describe a statistical application to the
scoring of sequential forecast distributions which provoked the
discovery.
\end{abstract}

\end{frontmatter}

\section{Scope, Motivation
 and Background}\label{s1}

The \emph{entropy} measure of a probability distribution has had
a myriad of useful applications in information sciences since its
full-blown introduction in the extensive article of \citet{Shan48}.
Prefigured by its usage in thermodynamics by Boltzmann and Gibbs,
entropy has subsequently bloomed as a showpiece in theories of
communication, coding, probability and statistics. So widespread is
its application and advocacy, it is surprising to realize that this
measure has a complementary dual which merits recognition and
comparison, perhaps in many realms of its current application, a
measure we term \emph{extropy}. In this article we display several
intriguing properties of this information measure, resolving a
fundamental question that has surrounded Shannon's measure since its
very inception.  The results provide links to other notable
information functions whose relation to entropy have not been
recognized. In particular, the standard $L_2$ distance between two
densities is identified as dual to the entropic measure of
Kullback--Leibler, an understanding provoked by considering the
extropy function as a Bregman function. We shall follow Shannon's
original notation and extend~it.

If $X$ is an unknown but
observable quantity with a finite discrete range of possible values
$\{x_1, x_2, \ldots, x_N\}$ and a probability mass function (p.m.f.)
vector $\mathbf{p}_N = (p_1, p_2, \ldots, p_N)$, the Shannon entropy
measure denoted by $H(X)$ or $H(\mathbf{p}_N)$ equals $-\sum_{i=1}^N
p_i \log(p_i)$. Its complementary dual, to be denoted  by $J(X)$ or
$J(\mathbf{p}_N)$, equals $ - \sum_{i=1}^N  (1-p_i) \log(1-p_i)$.
We propose this as the measure of \mbox{extropy}. As is entropy, \mbox{extropy} is
interpreted as a measure of the amount of uncertainty represented by
the distribution for $X$. The duality of $H(\mathbf{p}_N)$ and $J(\mathbf{p}_N)$ will be found to derive formally from the symmetric
relationship they bear with the sums of the (entropies, extropies)
in the $N$ crude event partitions defined by $[(X = x_i), (X \neq
x_i)]$. The complementarity of $H$ and $J$ arises from the fact that
the extropy of a mass function, $J(\mathbf{p}_N)$, equals a location
and scale transform of the entropy of another mass function that is
complementary to $\mathbf{p}_N$: that is,
\[
J(\mathbf{p}_N) = (N-1)\bigl[H(\mathbf{q}_N) - \log(N-1)
\bigr] ,
\]
 where
$\mathbf{q}_N
= (N-1)^{-1}(\mathbf{ 1}_N - \mathbf{p}_N)$.  This p.m.f. $\mathbf{q}_N$ is
constructed by norming the probabilities of the events $\tilde{E}_1,
\ldots, \tilde{E}_N$ which are complementary to $E_1, \ldots, E_N$. When
$N=2$ this yields the standard p.m.f. for $\tilde{E}_1$ as opposed to
the p.m.f. for $E_1$. Together, these two relationships establish
extropy as the complementary dual of entropy.

In his seminal
article that characterized the entropy function,
\citeauthor{Shan48} (\citeyear{Shan48}) began by formulating three
properties that might well be required of \emph{any} function $H(\cdot)$
that is meant to measure the amount of information inhering in a p.m.f.
$\mathbf{p}_N$. He
suggested  the following three properties as axioms for $H(\mathbf{p}_N)$:
\begin{enumerate}[(iii)]
 \item[(i)]  $H(p_1, p_2, \ldots, p_N)$ is continuous in
every argument;

\item[(ii)]  $H(\frac{1}{N}, \frac{1}{N}, \ldots,
\frac{1}{N})$ is
a monotonic increasing function of the dimension $N$;   and

\item[(iii)] for any positive integer $N$, and any values of
$p_i$ and $t$ each in $[0,1]$,
\begin{eqnarray*}
&&H\bigl(p_1,\ldots,p_{i-1},tp_i,
(1-t)p_i,p_{i+1},\ldots,p_N\bigr)
\\
&&\quad= H(p_1, p_2, \ldots, p_N) +
p_i H(t,1-t).
\end{eqnarray*}
\end{enumerate}
   Shannon then proved that the entropy function $H(\mathbf{p}_N) = -\sum_{i=1}^N p_i \log(p_i)$ is the \emph{only} function of
$\mathbf{p}_N$ that satisfies these axioms. It is unique up to an
arbitrary specification of location and scale.  Subsequently, the
article of \citet{Reny61} presented alternative characterizations of
entropy due to Fadeev and himself. These involved alternating these
axioms with various properties of Shannon's function, such as its
invariance with respect to permutations of its arguments and its
achieved
maximum occurring at the uniform distribution.

Shannon's third axiom concerns the behavior of the function $H(\cdot)$
when any category of outcome for $X$ is split into two
distinguishable possibilities, and the probability mass function
$\mathbf{p}_N$ is thereby refined into a p.m.f. over $(N+1)$
possibilities. It implies that the entropy in a joint distribution
for two quantities equals the entropy in the marginal distribution
for one of them plus the expectation for the entropy in the
conditional distribution for the second given the first:
%
\begin{eqnarray}
\label{eq:entropyjoint} \quad H(X,Y) &=& H(X) \nonumber\\[-8pt]\\[-8pt]
&&{}+ \sum_{i=1}^N
P(X=x_i) H(Y|X = x_i) .\nonumber
\end{eqnarray}
 The appeal of this result was a motivation favoring
Shannon's choice of his axiom (iii). However, in his original article
Shannon  slighted his own characterization theorem for entropy,
noting in a discussion (page~393) that its motivation is unclear
and that it is in no way necessary for the larger theory of
communication he was developing. He viewed it merely as lending
plausibility to some subsequent definitions. He considered the real
justification of the three axioms for entropy to reside in the
useful applications they support. In particular, he regarded the
implication of equation~(\ref{eq:entropyjoint}) as welcome
substantiation for considering $H(\cdot)$ as a reasonable measure of
information.

While the relevance of entropy to a wide array of
important applications has emerged over the subsequent half-century,
Shannon's attitude toward the foundational basis for entropy has
persisted. As one important example, the synthetic exposition of
\citet{CoTh91} begins directly with now common definitions required
for further developments and analysis, along with an unmotivated
specification of the entropy axioms. The authors found it
``irresistible to play with their relationships and interpretations,
taking faith in their later utility'' (page~12). They did so with
flair, exposing various roles understood for entropy in the fields
of electrical engineering, computer science, physics, mathematics,
economics and philosophy of science. In a similar vein, the
stimulating published lectures of \citet{Cati12} reassert and clarify
this standard take on axiomatic issues.  Caticha writes (page~79)
that ``both Shannon and Jaynes agree that one should not place too
much significance on the axiomatic derivation of the entropy
equation, that its use can be fully justified a-posteriori by its
formal properties, for example by the various inequalities it
\mbox{satisfies}. Thus, the standard practice is to define `information' as
a technical term using the entropy equation and proceed.  Whether
this meaning is in agreement with our colloquial meaning is another
issue. \ldots the difference is not about the equations but about what
they mean, and ultimately, about how they should be used.''  Caticha
considers such issues in his development of a conceptual
understanding of physical theory.

Forthrightly, the thoughtful
discussion of \citeauthor{Jayn03} [(\citeyear{Jayn03}), Section~11.3]  explicitly recognized and
addressed the discussable open status of Shannon's third axiom
characterizing entropy. Should this axiom really be \emph{required}
of any measure of the amount of uncertainty in a distribution?
Despite recognizing its crucial role in specifying Shannon's entropy
function mathematically, Jaynes was not convinced that an adequate
foundation for the \emph{uniqueness} claims of entropy as an
information measure had been found.  He concluded this long section
of his book by writing (\cite[page~351]{Jayn03}) ``Although the above
demonstration appears satisfactory mathematically, it is not yet in
completely satisfactory form conceptually.  The functional equation
(Shannon's third axiom) does not seem quite so intuitively
compelling as our previous ones did. In this case, the trouble is
probably that we have not yet learned how to verbalize the argument
leading to [axiom (iii)] in a fully convincing manner. Perhaps this
will inspire others to try their hand at improving the verbiage that
we used just before writing [axiom (iii)].''

In fact, Jaynes
appended an ``Exercise 11.1'' to his discussion, concluding with an
injunction to ``Carry out some new research in this field by
investigating this matter; try either to find a possible form of the
new functional equations, or to explain why this cannot be done.''
Concerns with claims regarding the uniqueness of entropy (along with
other matters regarding continuous distributions which we shall
address in this article) had also been aired by
\citeauthor{Kolm56} (\citeyear{Kolm56}), page~105.

Nonetheless, Jaynes clearly expected
that a satisfactory motivation for the special status of entropy as
a measure of information would be found, thinking that his
``exercise'' would be resolved with a solution explaining ``why this
cannot be done.'' In a direct sense, our construction and analysis
of the extropy measure shows the exercise to be solved rather by an
exhibition of the long sought ``new functional equation.''  We shall
specify this in our Result 3, which provides an alternative to
Shannon's third axiom and yields a different information measure.
The results of the present article show that the extropy measure,
far from generating inconsistencies which Jaynes feared (page~350), is
actually a complementary dual of the entropy function. The two
measures are clearly distinct, yet are fundamentally intertwined
with each other.  In tandem with Shannon's entropy measure denoted
by $H(\cdot)$, we respectfully denote our extropy measure by $J(\cdot)$.
It provides a resolution to Jaynes' insightful concerns and
accomplishments.

Our recognition of extropy as the
complementary dual of entropy emerged from a critical analysis and
completion of the logarithmic scoring rule for distributions in
applied statistics. Proper scoring rules are functions of forecast
distributions and the realized observations of the quantities at
issue. According to the subjectivist understanding of probability
and statistics as promoted by Bruno de Finetti, the assessment of
proper scoring rules for proposed forecasting distributions replaces
the role of hypothesis testing in objectivist methods.  None of an
array of proposed probability distributions can be considered to be
right or wrong. Each merely represents a different point of view
regarding a sequence or collection of unknown but observable
quantities.  The applied assessment of proper scoring rules provides
a method for evaluating the comparative qualities of the competing
points of view in the face of actual observed values of the
quantities as they come to be known.  The scoring functions are
intimately related to the theory of utility.  Such rules can also be
used to aid in the elicitation of subjective probabilities.

The so-called logarithmic score has long been touted for its
uniqueness in a specific respect relative to other proper scoring
rules.  The application we shall introduce raises issues concerning
its incompleteness in assessing asserted distributions. We shall
discuss details after the analysis of the duality of entropy and
extropy is exposed.  It will then be clear that the expected
logarithmic score of a distribution $\mathbf{p}_N$ coincides with
$-H(\mathbf{p}_N)$, which is called negentropy. The \emph{completion} of
the log score, which is motivated for a specific application,
involves the assessment of negextropy as well.

\begin{figure*}[t]

\includegraphics{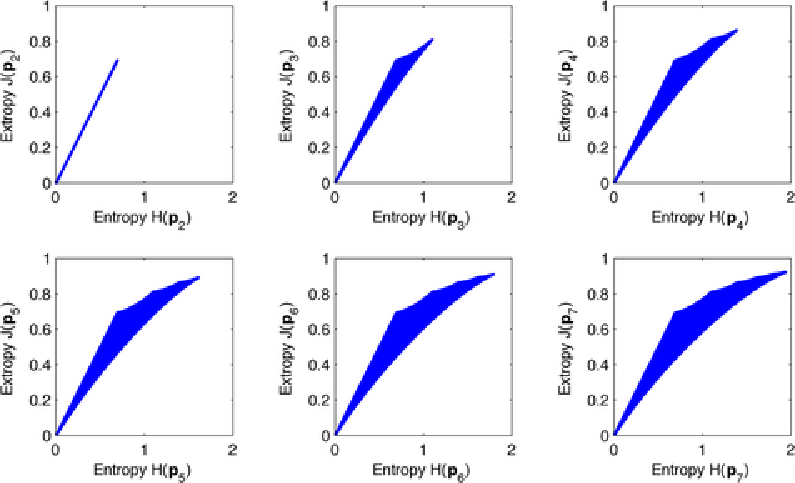}

\caption{The range of (entropy, extropy) pairs $(H(\cdot),J(\cdot))$
corresponding to all distributions within the unit-simplex of
dimensions $1$ through~$6$.  The ranges of the quantities they
assess have sizes $N=2$ through $7$.} \label{fig:ent2to7startYES}
\end{figure*}

After developing
the formal dual structure of the paired (entropy, extropy) functions
in Sections~\ref{s2}--\ref{s5} of this article, we shall outline
in Section~\ref{s6} the
role that extropy plays in the scoring of forecasting distributions,
using the \emph{Total} log scoring rule. We present the
axiomatization of extropy relative to entropy in Section~\ref{s2}, focusing
on an alternative to axiom (iii). In~Section~\ref{s3} we display graphically
the contours of the dual measures for the case of $N=3$. Section~\ref{s4}
identifies the dual equations and the complementary contraction
mapping.  In Section~\ref{s5} we develop the theory for continuous density
functions, formalizing differential and relative (entropy, extropy)
in the context of general Bregman functions.  We show how relative
extropy arises as a second directed distance function that is a
complementary dual to the Kullback--Leibler divergence, the standard
formulation of relative entropy. Section~\ref{s7} presents a concluding
discussion.

\section{The Characterization of Extropy}\label{s2}

  \textit{Context}:  Consider an observable quantity $X$ with
possible values contained in the range \textsl{R}$(X) = \{x_1, x_2,
\ldots, x_N\}$. The vector $\mathbf{p}_N = (p_1, p_2, \ldots, p_N)$ is
composed of probability mass function values asserted for $X$ over
the event partition $[(X = x_1), (X = x_2), \ldots, (X = x_N)]$. Though
we typically refer to $\mathbf{p}_N$ as a p.m.f., we sometimes use common
parlance that is an abuse of \emph{formal} terminology, referring to
it as a
``distribution.''  To begin our discussion, we recall the following:

\begin{definition}\label{d1}
The \emph{entropy} in $X$ or in $\mathbf{p}_N$ equals
%
\begin{equation}
H(X) = H(\mathbf{p}_N) \equiv - \sum_{i=1}^N
p_i \log(p_i) .
\end{equation}
\end{definition}

We note
that we use natural logarithms as opposed to base 2, and we
introduce the following:

\begin{definition}\label{d2}
The \emph{extropy} in $X$ or in $\mathbf{p}_N$ equals
%
\begin{equation}
\quad \quad J(X) = J(\mathbf{p}_N) \equiv - \sum_{i=1}^N
(1-p_i) \log(1-p_i) .
\end{equation}
\end{definition}

\begin{result}\label{r1}
If $N=2$, so $X$ denotes merely an event,
then $H(X) = J(X)$, but when $N \geq 3,   H(\mathbf{p}_N) > J(\mathbf{p}_N) $ as long as $\mathbf{p}_N$ contains three or more positive
components.
\end{result}

Clearly, $H(\mathbf{p}_2)  =  -p_1
\log(p_1) - (1-p_1) \log(1-p_1) =  J(\mathbf{p}_2)$.  An algebraic
proof of Result \ref{r1} appears in Appendix \ref{appA}.  However, its truth is
apparent easily from computational examples.
Figure~\ref{fig:ent2to7startYES} displays the range of possibilities
for the (entropy, extropy) pairs for probability mass functions
within the unit-simplexes of dimensions $1$ through $6$ (values of $N
= 2$ through $7$).

Evidently, the range of possible (entropy, extropy) pairs
at each successive value of $N$ incorporates the range for the
previous value of $N$, with another section merely attached to this
range. Notice particularly that the range of possible (entropy,
extropy) pairs is \emph{not} convex. As viewed across the six
examples shown in Figure~\ref{fig:ent2to7startYES}, the range
exhibits convex scallops along its upper boundary:  there are
$(N-2)$ scallops and one flat edge along its upper boundary for the
unit-simplex of dimension $(N-1)$. The flat edge as the northwest
boundary is the line defined by $H(p,1-p) = J(p,1-p)$, running in
the southwest to northeast direction from $(0,0)$ to $(-\log(0.5),
-\log(0.5))$. The lower boundary of the range of pairs is a single
concave scallop, ruling its own
interior out of the range of possible (entropy, extropy) pairs.

\begin{result}\label{r2}
 $J(X)$ satisfies Shannon's axioms (i) and
(ii).
\end{result}

 The function $J(\cdot)$ is evidently continuous in its
arguments [axiom (i)], and
 \begin{eqnarray*}
J\biggl(\frac{1}{N}, \frac{1}{N}, \ldots, \frac{1}{N}\biggr)
&=&-N \biggl(1-\frac{1}{N}\biggr) \log\biggl(1-\frac{1}{N}\biggr)
\\
&=& (N-1)\bigl[\log(N)-\log(N-1)\bigr]
\end{eqnarray*}
is a monotonic increasing function of $N$ [axiom (ii)].

\subsection{Further Shared Properties of \texorpdfstring{$H(\cdot)$}{$H(cdot)$} and \texorpdfstring{$J(\cdot)$}{$J(cdot)$}}

As to other touted properties of entropy, extropy shares many of
them.  For example, the extropy measure is obviously permutation
invariant. It is also invariant with respect to monotonic
transformations of the variable $X$ into $Y = g(X)$. Moreover, for
any size of $N$, the maximum extropy distribution is the uniform
distribution. This can be proved by standard methods of constrained
maximization using Lagrange multipliers.  Let $L(\mathbf{p}_N,\lambda)$
be the Lagrangian expression for the extropy of $\mathbf{p}_N$ subject
to the constraint $\sum p_i = 1 $:
\[
L(\mathbf{p}_N,\lambda) = -\sum_{i=1}^N
(1-p_i) \log(1-p_i) + \lambda \Biggl(1 - \sum
_{i=1}^N p_i\Biggr) .
\]
The $N$ partial derivatives have the form $\frac{\partial
L}{\partial p_i}   =   \log  (1-p_i)
 +  1  -  \lambda$.  Setting each of these equal to $0$ yields $N$
 equations of the form $\lambda = 1 + \log  (1-p_i)$.  These $N$ equations,
 together with $\frac{\partial L}{\partial \lambda} = 0$, ensure that
 all the $p_i$ are equal, and thus they must each equal $1/N$.
 Second order conditions for a maximum are satisfied at this first order solution.
 Analysis of the boundaries of the unit-simplex constraining
 $\mathbf{p}_N$ yields the
 \emph{minimum} values of extropy at the vertices:
$J(\mathbf{
 e}_i) = 0$ for each echelon basis $\mathbf{
 e}_i \equiv (0,0, \ldots, 0, 1_i, 0, \ldots, 0)$ with $i = 1, 2, \ldots, N$.

As to differences in the two measures, notice that the scale of the
maximum entropy measure is unbounded as $N$ increases, because
$H(\frac{1}{N}, \frac{1}{N}, \ldots, \frac{1}{N})  =  \log(N)$.  In
contrast, the scale of the maximum extropy is bounded by 1, for
$J(\frac{1}{N}, \frac{1}{N}, \ldots, \frac{1}{N})  =
(N-1) \log[N/(N-1)]$.  The limit of 1 can be determined by
recognizing that
\begin{eqnarray*}
&&\lim_{N \rightarrow \infty}(N-1) \log \biggl(\frac{N}{N-1} \biggr)
\\
&&\quad= \lim_{N \rightarrow \infty}\log \biggl(1+\frac{1}{N-1}
\biggr)^
{N-1} = \log(e) = 1 .
\end{eqnarray*}

\subsection{The Extropy Measure of a Refined Distribution}

We can now examine precisely how and why extropy does \emph{not}
satisfy Shannon's third axiom for entropy, and how it does behave
with respect to measuring the refinement of a probability
distribution. Algebraically, the refinement axiom for extropy arises
from its definition, which yields the following result:

\begin{result}\label{r3}
 For any positive integer $N$, and any
values of
$p_i$ and $t$ each in $[0,1]$,
\begin{eqnarray*}
&& J\bigl(p_1,\ldots,p_{i-1},tp_i,
(1-t)p_i,p_{i+1},\ldots,p_N\bigr)
\\
&&\quad = J(p_1, p_2, \ldots, p_N) +
\triangle(p_i,t) ,
\end{eqnarray*}
 where
\begin{eqnarray*}
\triangle(p_i,t) &=& (1-p_i) \log(1-p_i)
\\
&&{}- (1-tp_i) \log(1-tp_i)
\\
&&{}-\bigl[1-(1-t) p_i\bigr]  \log\bigl[1-(1-t) p_i
\bigr] .
\end{eqnarray*}
\end{result}

This follows directly from the definition of $J(\mathbf{p}_N)$.  The structure of the gain to a refined extropy,
$\triangle(p_i,t)$, can be recognized by introducing a function
$\varphi(p) \equiv (1-p)\log(1-p)$ and noting that $\triangle(p_i,t)
=\varphi (p_i) - [\varphi (tp_i) + \varphi ( (1-t)p_i )]$. This
difference can be shown to be always nonnegative.

\begin{figure*}

\includegraphics{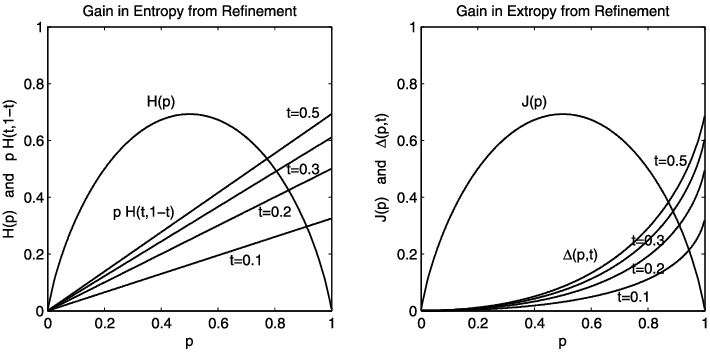}

\caption{Entropy and extropy for a refined distribution $[tp,
(1-t)p, 1-p]$ both equal the entropy or extropy for the base
probabilities $(p, 1-p)$ plus an additional component.}
\label{fig:extentropyrefinedfig}
\end{figure*}

 Result \ref{r3} is easily interpreted visually when $N = 2$. The
left panel of Figure~\ref{fig:extentropyrefinedfig} displays the
difference between the entropies $H(tp, (1-t)p, 1-p)$ and $H(p,1-p)$
according to Shannon's axiom (iii).  The right panel displays the
extropy $J(p,1-p)$ along with the difference between the extropies
$J(tp, (1-t)p, 1-p)$  and  $J(p,1-p)$ according to Result \ref{r3}.  The
important feature of the display is the difference between
$pH(t,1-t)$ on the left and $\triangle(p,t)$ on the right, a
difference which does not depend on the magnitude of $N$.  In each
panel, the differences are shown as functions of $p \in [0,1]$ for
the four values of $t  =  0.1, 0.2, 0.3$ and $0.5$. For any value of
$t$, the difference functions $\triangle(p,t) = \triangle(p,t')$ for
$t' = (1-t)$.

According to Shannon's axiom (iii), the
entropy for the refined mass function $[tp, (1-t)p, 1-p]$ increases
linearly with $p$ at the rate of the entropy in the refining split
factor, $H(t,1-t)$.  In contrast, the extropy of the refined
distribution increases at an increasing rate as a function of $p$.
For small values of $p$, the extropy of the refined distributions
increases more slowly with $p$ than does entropy, while for large
values of $p$ it increases more quickly. When the value of $p$
equals 1, the values of the entropy and extropy of the refined
distribution equalize, for each $t \in [0,1]$. This results from the
fact that when $p=1$, the refined distribution is virtually a binary
distribution $(t, 1-t, 0)$, for which entropy and extropy are equal.
In this case the distribution being refined would be a degenerate
distribution representing certainty.

As a gauge of the
increase in uncertainty provided when a distribution is refined,
this nonlinear feature of the extropy measure is appealing in its
own right. Refining a larger probability with a splitting factor of
size $t$ may well be considered to increase the amount of
uncertainty that is specified at a greater rate than when refining a
smaller probability by this same factor. Consider two ways of
refining a mass function $\mathbf{p}_2 = (0.04, 0.96)$, for example, into
$\mathbf{p}_3 = (0.01, 0.03, 0.96)$ as opposed to $\mathbf{p}_3 =
(0.04, 0.24,
0.72)$.  In both cases, one of the probabilities is refined into two
pieces in the ratio of $1\dvtx 3$. Examine the values of $\triangle(0.04,
0.25)$ and $\triangle(0.96, 0.25)$ in Figure~\ref{fig:extentropyrefinedfig}(right). Although the
rate of increase in entropy due to the refinement of either
probability $p_i$ is identical in the two cases, the rate of
increase in extropy when refining the component $p_i=0.04$ is nearly
zero, while it is far greater when refining the larger probability
component $p_i = 0.96$. It is a natural feature of the extropy
function that this information measure adjusts toward the maximum
entropy/extropy more quickly the more quickly the refined
distribution adjusts toward the uniform.

\begin{figure*}

\includegraphics{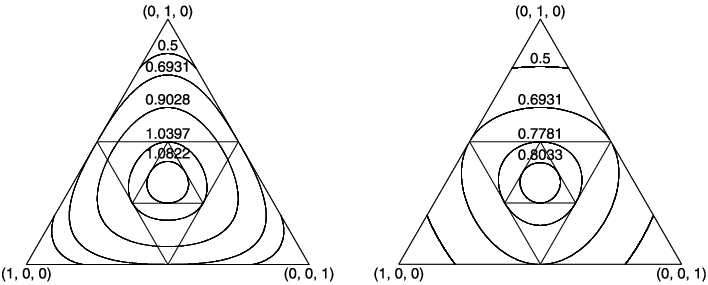}

\caption{At left are contours of equal entropy distributions within
the 2-D unit-simplex, $\mathbf{S}^2$.  At right are contours of equal
extropy distributions. The relevance of the inscribed triangles
shall become
apparent in Section~\protect\ref{s4}.}
 \label{fig:entandextcontours}
\end{figure*}

Replacing Shannon's axiom (iii) with our Result \ref{r3} would complete
an axiomatic characterization of extropy.  When $N=1$, the
specifications of axiom (iii) and Result \ref{r3} are algebraically
identical, yielding $H(t, 1-t) = J(t,1-t)$.  When $N \geq 2$ the
bifurcation first occurs. In this context, Result \ref{r3} can then be seen
to be a generator of the entire function $J(\mathbf{p}_N)$ for all
values of $N$. The extropy function is the unique function that
adheres to Shannon's axioms (i) and (ii) \emph{and} to the content of
Result \ref{r3}, considered as an axiom.

\section{Isoentropy, Isoextropy Contours in the Unit-Simplex}\label{s3}

For the graphical displays that follow, we suppose that a quantity
$X$ has range $ R (X) = \{1,2,3\}$ and that these
possibilities are assessed with a probability mass function
$\mathbf{p}_3$ in the unit-simplex $\mathbf{S}^2$.
 Figure~\ref{fig:entandextcontours}(left) displays some contours of constant
entropy distributions in the 2-dimensional unit-simplex $(N=3)$ to
compare with some contours of constant extropy distributions in
Figure~\ref{fig:entandextcontours}(right). These contours exhibit a
geometrical sense in which the extropy and entropy measures of a
distribution are complementary. Whereas entropy contours sharpen
into the vertices of the simplex and flatten along the faces, the
extropy contours sharpen into the midpoints of the faces and flatten
toward the vertices.

 Further understanding can be gained from Appendix~\ref{appB} which
displays the single isoentropy contour at $H(\mathbf{p}_3) = 0.9028$
along with some members of the range of isoextropy contours that
intersect with it.  A~computable application in astronomy is
mentioned.

\section{Extropy as the Complementary Dual of~Entropy}\label{s4}

Two behaviors identify the mathematical relation of extropy to
entropy as its complementary dual.  To begin, the duality is
distinguished by a pair of symmetric equations relating the sum of
the entropy and extropy of a distribution to the
entropies and extropies of their component probabilities.

\begin{result}\label{r4}
\begin{eqnarray*}
H( \mathbf{p} _N) + J( \mathbf{p} _N) &= & \sum
_{i=1}^N H(p_i,1-p_i)
\\
& =& \sum_{i=1}^N J(p_i,1-p_i).
\end{eqnarray*}
\end{result}

 This equation for the sum of $H(\mathbf{p}_N)$ and $J(\mathbf{p}_N)$ derives from summing separately the two components of each
$H(p_i, 1-p_i)  =  - p_i \log(p_i) - (1-p_i) \log(1-p_i)  =
J(p_i, 1-p_i)$ over values of $i = 1, 2, \ldots, N$.  This simple
result identifies the symmetric dual equations that relate extropy
to entropy:
\[
J(\mathbf{p}_N) = \sum_{i=1}^N
H(p_i,1-p_i) - H(\mathbf{p}_N),
\]
and symmetrically,
\[
H(\mathbf{p}_N) = \sum_{i=1}^N
J(p_i,1-p_i) - J(\mathbf{p}_N).
\]
These two equations, symmetric in $H(\cdot)$ and $J(\cdot)$, display that
the extropy of a distribution equals the difference between the sum
of the entropies over the \mbox{crudest} partitions defined by the possible
values of $X$, that is, $[(X=x_i), (X \neq x_i)]$, and the entropy in
the finest partition they define,  $[(X=x_1), (X=x_2), \ldots ,
(X=x_N)]$. Extropy and entropy can each be represented by the same
function of the other. Since these two functions differ only in the
refinement axioms that generate them, it is apparent that their
symmetric duality is fundamentally related to the refinement
characteristics inherent in
their third axioms.

\begin{figure*}[b]

\includegraphics{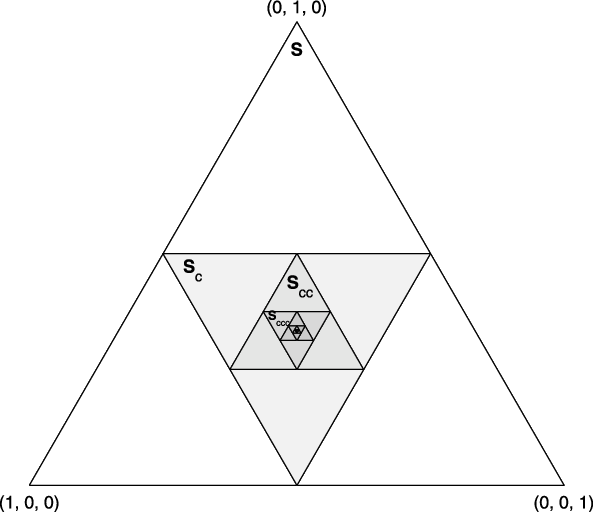}

\caption{The complementary
distribution
mapping contracts the unit-simplex $\mathbf{S}$ into the inscribed simplex
$\mathbf{S}_c$, which it contracts in turn into the inscribed $\mathbf{S}_{cc}$,
and then into $\mathbf{S}_{ccc}$ and so on.}
\label{fig:contractionsimplexes}
\end{figure*}

As to the complementarity of their relation, it is based on
generalizing the notion of a complementary event to a complementary
quantity.  Relative to a probability mass function $\mathbf{p}_N$ for a
partition vector $[{(X=x_1),} (X=x_2), \ldots , (X=x_N)]$, \emph{define
the complementary mass function} as $\mathbf{q}_N = (N-1)^{-1}(\mathbf{
1}_N - \mathbf{p}_N)$.  The general complementary mass function $\mathbf{q}_N$ can be considered to specify a ``distribution of
unlikeliness'' of the possible values of $X$, as opposed to $\mathbf{p}_N$ which distributes the assessed likeliness of the possible
values. If $N=2$, complementarity specifies $\mathbf{q}_2 = (q_1, q_2)
= (1-p_1, 1-p_2) = (p_2, p_1)$. This merely identifies the
arbitrariness of analyzing an event in terms of $E_1$ and its
complement $\tilde{E}_1 = E_2$,\vspace*{1pt} as opposed to $F_1 = \tilde{E}_1$
and \emph{its} complement $\tilde{F}_1 = E_1$.  For larger values of
$N$, however, general complementarity generates $\mathbf{q}_N$ from $\mathbf{p}_N$ as a truly distinct mass function. In these terms, the
general relation between $H$ and $J$ is that the extropy of a p.m.f.
$\mathbf{p}_N$ equals a linearly rescaled measure of entropy of its
complementary p.m.f. $\mathbf{q}_N$.

\begin{result}\label{R5}
\[
J(\mathbf{p}_N) = (N-1) \bigl[H(\mathbf{q}_N) -
\log(N-1)\bigr] .
\]
\end{result}

 To be explicitly clear, the extropy of $\mathbf{p}_N$ is \emph{not} a rescaled value of the entropy of $\mathbf{p}_N$.  It \emph{is} a
rescaled value of the entropy of the general complement of $\mathbf{p}_N$.

 This result follows from simple algebra.  Structurally, the
entropy measure of a probability mass function has a complementary
dual in its extropy measure, which derives from the entropy of a
complementary mass function.  In turn, this complementary mass
function has its own extropy.  However, this extropy value does \emph{not} derive from the entropy of the original p.m.f., but from a further
\emph{complement of this complement}.

Most statisticians will
be familiar with the notion of duality from the fact that any linear
programming problem has a dual formulation in which the coefficient
vector of the linear objective function has a dual relation with the
vector of constraint values. The linear programming duality has the
feature that the dual structure of a dual problem yields the
original problem structure.  Duals with this property are called
``involutions.'' As we shall see now, the duality of extropy with
entropy does \emph{not} prescribe an involution, but rather a second
distinct structure.

The mapping of a probability mass function
$\mathbf{p}_N$ to its complement $\mathbf{q}_N = (N-1)^{-1} (\mathbf{1}_N -
\mathbf{p}_N)$ is a contraction mapping. Every mass function in a
unit-simplex is mapped onto a complementary function lying within an
inscribed simplex of the same dimension. In turn, this complementary
mass function has its own complementary distribution lying within a
simplex inscribed in that one.  The fixed-point theorem for
contraction mappings assures that the uniform distribution in the
center of the unit-simplex is the unique mass function whose
complementary mass function equals itself. Figure~\ref{fig:contractionsimplexes} displays the way this contraction
works in two dimensions for mass functions $\mathbf{p}_3$. Notice that
the points in the vertex triangles of the unit-simplex are not
contraction images of any other point in the unit-simplex. Thus, the
formal complementary duality of $H(\cdot)$ and $J(\cdot)$ with respect to
$\mathbf{p}_N$ and $\mathbf{q}_N$ inheres in their forward and backward
images rather than a cyclic image. The dual is not an
involution.

A numerical example detailing how the isocontours of
$H(\cdot)$ generate isocontours of $J(\cdot)$ appears in Appendix~\ref{appC}.

\section{Differential Extropy and Relative Extropy
 for Continuous Distributions}\label{s5}

Devising the extropy measure of a continuous distribution admitting
a density function yields a pleasant surprise. As to entropy,
\citeauthor{Shan48} [(\citeyear{Shan48}), page~628]
  had initially proposed that the entropy
measure $-\sum p_i \log(p_i)$ has an analogue in the definition
$-\int f(x) \log f(x) \,dx$ when the distribution function for a
variable $X$ admits a continuous density. He motivated this (page~623) by the idea that refining the categories for a discrete
quantity $X$, with diminishing probabilities in each, yields this
analogous definition in the limit. This definition has subsequently
become known as ``differential entropy.'' In a critical and
constructive review, \citet{Kolm56} concurred with Shannon's
suggestion, but with qualifying reservations regarding its
noninvariance with respect to monotonic transformations of the
variable $X$ and its relativity to a uniform dominating measure
over the domain of $X$. His clarifications established a more
general definition of ``relative entropy'' which includes
differential entropy as a special case. Relative entropy was
analyzed in measure theoretic detail in the classic work of
\citet{Kull59}.  Now known as the Kullback--Leibler divergence (or
directed distance) between a density $f(\cdot)$ and a related
absolutely continuous density $g(\cdot)$, this is defined for the
continuous case as $D(f\|g) \equiv \int f(x) \log \frac{f(x)}{g(x)}
\,dx$. When $g(x)$ is the special case of a uniform density, this
reduces to Shannon's
definition of differential entropy.

The dual complementarity of extropy with entropy for continuous
densities can be derived in the context of relative entropy. The
details are couched in the language of general Bregman functions,
which unifies the discrete theory as well. We shall develop these
results forthwith. For a novice reader of these ideas, the
development of continuous \emph{differential} entropy and extropy in
the style suggested by Shannon is perhaps more instructive.  It
motivates the definition of differential extropy as $- \frac{1}{2}
\int  f^2(x)\, dx$.  The role played by the uniform dominating
measure in generating this integral will be apparent. We present an
introductory analysis in Appendix \ref{appD}. We now begin directly by
developing the more general formulation of relative extropy as the
dual to relative entropy in a discrete context, and then pursuing
the continuous
analysis using Bregman functions.

\subsection{(Relative Entropy, Relative Extropy) for Two Mass
Functions:  Kullback's Directed Distance and Its Complementary
Dual}\label{s5.1}

We continue to work in the context of a considered quantity whose
possible values generate the finite partition vector $[(X=x_1),
(X=x_2), \ldots, (X=x_N)]$. Suppose that the vector $\mathbf{s}_N$
represents a second
p.m.f., distinct from $\mathbf{p}_N$.  In this context
we recall the following:

\begin{definition}\label{d3}
The \emph{relative entropy} of $\mathbf{p}_N$ with respect to
$\mathbf{s}_N$ is defined as the
Kullback--Leibler divergence to equal
%
\begin{equation}
\label{eq:DefDps} D(\mathbf{p}_N \| \mathbf{s}_N) \equiv
\sum_{i=1}^N p_i \log\biggl(
\frac{p_i}{s_i}\biggr) .
\end{equation}
\end{definition}

 Notice that this definition does \emph{not} involve a minus
sign in front, as  $D(\mathbf{p}_N \| \mathbf{s}_N)$ is always
nonnegative. It makes no difference whether the variable $X$ is
transformed by any monotone function to a new variable $Y$:  the
relative entropy in $\mathbf{p}_N$ with respect to $\mathbf{s}_N$ remains
the same. We recall that this \emph{directed} distance function is
not symmetric in $\mathbf{p}_N$ and $\mathbf{s}_N$, and thus its name.

To define the relative extropy of $\mathbf{p}_N$ with respect
to
$\mathbf{s}_N$, we follow the same tack as in defining extropy
\mbox{itself}:

\begin{definition}\label{d4}
The \emph{relative extropy} of $\mathbf{p}_N$ with respect
to $\mathbf{s}_N$ is defined by a function
complementary to the Kullback--Leibler divergence as
\[
D^c(\mathbf{p}_N \| \mathbf{s}_N) \equiv
\sum_{i=1}^N (1-p_i) \log
\biggl(\frac{1-p_i}{1-s_i}\biggr) .
\]
\end{definition}

\begin{result}\label{r6}
When the p.m.f. $\mathbf{s}_N$ happens to be
the uniform p.m.f. $\mathbf{u}_N = N^{-1}\mathbf{1}_N$, the relative entropy
and extropy measures return to rescaled values of the discrete
entropy and extropy measures with which we are familiar:
\begin{eqnarray*}
D(\mathbf{p}_N \|\mathbf{u}_N) &= & \sum
_{i=1}^Np_i \log\biggl(
\frac{p_i}{1/N}\biggr) = \log(N) - H(\mathbf{p}_N)
\\
&=& H(\mathbf{u}_N) - H(\mathbf{p}_N) ,
\end{eqnarray*}
and
\begin{eqnarray*}
 D^c(\mathbf{p}_N \|\mathbf{u}_N) &=& \sum
_{i=1}^N(1-p_i) \log \biggl(
\frac{1-p_i}{1-1/N} \biggr)
\\
& =& \sum_{i=1}^N(1-p_i)\log
\biggl(\frac{N}{N-1} \biggr)\\
&& {}+\sum_{i=1}^N(1-p_i)
\log(1-p_i)
\\
&  =& (N-1)\log \biggl(\frac{N}{N-1} \biggr) \\
&&{}+ \sum
_{i=1}^N(1-p_i) \log(1-p_i)
\\
& =& J(\mathbf{u}_N) - J(\mathbf{p}_N) .
\end{eqnarray*}
\end{result}

\subsubsection{The complementary equation}

It is straightforward to recognize that again, defining now two
complementary mass functions $\mathbf{q}_N = (N-1)^{-1}(\mathbf{1}_N -
\mathbf{p}_N)$ and $\mathbf{t}_N = (N-1)^{-1}(\mathbf{1}_N - \mathbf{s}_N)$, we
find that a complementary equation identifies $D^c(\mathbf{p}_N \| \mathbf{s}_N)$ as the K-L divergence between the p.m.f.'s complementary to
$\mathbf{p}_N$ and $\mathbf{s}_N$:

\begin{result}\label{r7}
\[
D^c(\mathbf{p}_N \|\mathbf{s}_N) = (N-1) D(
\mathbf{q}_N \| \mathbf{t}_N) .
\]
\end{result}

Moreover, an alternative algebraic manipulation of Definition \ref{d4}
provides that
%
\begin{eqnarray}
\label{eq:DcRepresentation} D^c(\mathbf{p}_N\|\mathbf{s}_N)
&=&\sum_{i=1}^N (1-p_i)
\log(1-p_i)
\nonumber
\\
&&{}-\sum_{i=1}^N (1-p_i+s_i-s_i)
\log(1-s_i)
\nonumber
\\
&=&\sum_{i=1}^N (1-p_i)
\log(1-p_i) \nonumber\\
&&{}-\sum_{i=1}^N
(1-s_i) \log(1-s_i)
 \\
&&{}+\sum_{i=1}^N (p_i-s_i)
\log(1-s_i)
\nonumber
\\
&=& J(\mathbf{s}_N)-J(\mathbf{p}_N)+ \sum
_{i=1}^N p_i \log \biggl(
\frac{1-s_i}{N-1} \biggr)
\nonumber
\\
&&{}-\sum_{i=1}^N s_i \log
\biggl(\frac{1-s_i}{N-1} \biggr) ,
\nonumber
\end{eqnarray}
 because $\sum_{i=1}^N   (p_i-s_i) \log(N-1)  =  0  $.
This yields another interesting and useful representation:

\begin{result}\label{r8}
\begin{eqnarray*}
D^c(\mathbf{p}_N \| \mathbf{s}_N) &= & J(
\mathbf{s}_N) - J(\mathbf{p}_N)
\\
&&{}+ E_{\mathbf{p}_N}\bigl[\log\bigl(t^o(X)\bigr)\bigr] -
E_{\mathbf{s}_N}\bigl[\log\bigl(t^o(X)\bigr)\bigr] ,
\end{eqnarray*}
where $t^o(X) \equiv \sum_{i=1}^N(X = x_i)t_i $.
\end{result}

 That is, $t^o(X)$ equals the component probability in the
$\mathbf{t}_N$ vector associated with the value of $X$ that happens to
be observed.  This holds algebraically because one of the event
indicators, $(X=x_i)$, equals 1 (since the equation it indicates is
true) while the other $(N-1)$ event indicators equal $0$.  The
equations they indicate are false.

The relative extropy value
of $\mathbf{p}_N$ relative to $\mathbf{s}_N$ equals the difference in
their extropy values, adjusted by a difference in two expectations
of a specific log mass function value:  the mass function component
of $\mathbf{t}_N$ associated with the particular partition event that
is found to occur. This is the mass function that is complementary
to $\mathbf{s}_N$. The usefulness of Result \ref{r8} shall arise as a
motivation for a definition of relative extropy between two
densities.

 The analogous result pertinent to the K-L divergence, deriving
from
(\ref{eq:DefDps}) would be as follows:\def\theResult{8$'$}

\begin{Result}\label{r8prime}
\begin{eqnarray*}
D(\mathbf{p}_N \| \mathbf{s}_N) &=& H(
\mathbf{s}_N) - H(\mathbf{p}_N)
\\
&&{}-E_{\mathbf{p}_N}\bigl[\log\bigl(s^o(X)\bigr)\bigr] +
E_{\mathbf{s}_N}\bigl[\log\bigl(s^o(X)\bigr)\bigr] ,
\end{eqnarray*}
 where $s^o(X) \equiv \sum_{i=1}^N(X = x_i)s_i $.
 \end{Result}

\subsubsection{Relative (entropy, extropy) of complementary mass functions}

A final note of interest concerns the pair of relative (entropy,
extropy) assessments between \emph{complementary} mass functions such
as $\mathbf{p}_N$ and $\mathbf{q}_N$.  The relative entropy of $\mathbf{p}_N$
with respect to $\mathbf{q}_N$ equals a translated expected value of
the asserted log odds ratio in favor of the occurring partition
event: $D(\mathbf{p}_N \| \mathbf{q}_N) = \sum_{i=1}^N  p_i
\log(\frac{p_i}{1-p_i}) + \log(N-1)$. Intriguingly, but again deriving
easily from a direct application of Definition \ref{d4}, their relative
extropy also equals $(N-1)$ times an expected log odds ratio in
favor of the occurring partition event too. However, \emph{this} odds
ratio is assessed in terms of the complementary distribution of
unlikeliness, $\mathbf{q}_N$, rather than in terms of the usual
distribution of likeliness, $\mathbf{p}_N$:
\begin{eqnarray*}
&&D^c(\mathbf{p}_N \|\mathbf{q}_N) \\
&&\quad= (N-1)
\Biggl[\sum_{i=1}^N q_i \log
\biggl(\frac{q_i}{1-q_i}\biggr) + \log(N-1)\Biggr].
\end{eqnarray*}
 Both of these interpretations as expected log odds ratios
are adjusted by an additive constant,\break $\log(N-1)$.  This additive
constant can be recognized as the\vspace*{2pt} expected log odds associated with
a uniform distribution: $ \sum_{i=1}^N  u_i
\log(\frac{u_i}{1-u_i}) = \sum_{i=1}^N  \abfrac{1}{N}\times\break
\log(\frac{\nfrac{1}{N}}{(1-\nfrac{1}{N})}) = -\log(N-1) $. Thus, we
have an interesting pair of representations for the
relative (entropy, extropy) between complementary mass functions:

\begin{result}\label{r9}
\begin{eqnarray*}
&&D(\mathbf{p}_N \|\mathbf{q}_N)
\\
&&\quad= E_{\mathbf{p}_N}\biggl[
\log\biggl(\frac{p^o}{1-p^o}\biggr)\biggr] - E_{\mathbf{u}_N}\biggl[\log\biggl(
\frac{u^o}{1-u^o}\biggr)\biggr] ,
\end{eqnarray*}
 and
\begin{eqnarray*}
 D^c(\mathbf{p}_N \|\mathbf{q}_N) &=& (N-1)
\biggl\{E_{\mathbf{q}_N}\biggl[\log\biggl(\frac{q^o}{1-q^o}\biggr)\biggr]\\
&&\hphantom{(N-1)
\biggl\{}{}-
E_{\mathbf{u}_N}\biggl[\log\biggl(\frac{u^o}{1-u^o}\biggr)\biggr]\biggr\} ,
\end{eqnarray*}
 where $p^o, q^o $ and $u^o$ are the probabilities
assessed for the value of $X$ that happens to be observed, as
assessed according to the p.m.f.'s $\mathbf{p}_N,
\mathbf{q}_N$  and $\mathbf{u}_N$, respectively.
\end{result}

\subsubsection{Unifying \texorpdfstring{$D(\cdot\|\cdot)$}{$D(cdot||cdot)$} and \texorpdfstring{$D^c(\cdot\|\cdot)$}{$D^c(cdot||cdot)$} as Bregman divergences}

The theory of Bregman functions both unifies our understanding of
the (entropy, extropy) duality and provides the basis for
formalizing their functional representations for continuous
densities. In this context it will yield still another surprise. The
text of \citet{CeZe97}  develops the general theory of
Bregman functions and a wide variety of applications.  In the
definition below we recall the notion of Bregman
divergence from \citet{BMDG05}:

\begin{definition}\label{d5}
Let $\mathcal{C}$ be a convex subset
of $\Re^N$  with a nonempty relative interior, denoted by
$\operatorname{ri}(\mathcal{C})$. Let $\Phi\dvtx \mathcal{C}\rightarrow \Re$ be a
strictly convex function, differentiable in $\operatorname{ri}(\mathcal{C})$. For
$\mathbf{p}_N, \mathbf{s}_N \in \mathcal{C}$ the \emph{Bregman divergence}
$d_{\Phi}\dvtx \mathcal{C}\times \operatorname{ri}(\mathcal{C}) \rightarrow \Re$
corresponding to $\Phi$ is given by
\begin{eqnarray*}
d_{\Phi}(\mathbf{p}_N, \mathbf{s}_N) &=& \Phi(
\mathbf{p}_N)-\Phi(\mathbf{s}_N)\\
&&{}-\bigl\langle \nabla\Phi(
\mathbf{s}_N), (\mathbf{p}_N-\mathbf{s}_N)
\bigr\rangle,
\end{eqnarray*}
 where
$\nabla\Phi(\mathbf{s}_N)$ is the gradient vector of  $\Phi$ evaluated
at $ \mathbf{s}_N$ and the angle brackets $\langle   \cdot, \cdot \rangle$ denote ``inner
product.''  The function $\Phi(\cdot)$ is called a \emph{Bregman function}.
\end{definition}

 An important special case of the Bregman function reduces
its action to the sum of a common function applied to each of the
components of a vector, that is, $\Phi(\mathbf{p}_N)=\sum_{i=1}^N\phi(p_i)$.  In this case the Bregman divergence
is said to be ``separable'' (\citeauthor{StVa12}, \citeyear{StVa12}), with the form
%
\begin{eqnarray}
\label{eq:separableBreg}&& d_{\Phi}(\mathbf{p}_N, \mathbf{s}_N)
\nonumber\\[-8pt]\\[-8pt]
&&\quad= \sum_{i=1}^N \bigl[\phi(p_i)-
\phi(s_i)-\phi'(s_i) (p_i-s_i)
\bigr] .\nonumber
\end{eqnarray}
A standard application of the separable case identifies the Shannon
entropy as a Bregman divergence.  Consider the component function
$\phi(p)=\varphi_1(p)$, where  $\varphi_1(p)\equiv p \log(p)$, which
identifies the vector Bregman function as $\Phi(\mathbf{p}_N) = -H(\mathbf{p}_N)$.  Since $\phi'(p) = \log(p) + 1$, a direct application of the
separable Bregman divergence form (\ref{eq:separableBreg}) yields
the following well-known result, which
is reported in \citet{BMDG05}:

\begin{result}\label{r10}
The Bregman divergence associated with
$\Phi(\mathbf{p}_N)=-H(\mathbf{p}_N)$ is
\[
d_{\Phi}(\mathbf{p}_N, \mathbf{s}_N) = \sum
_{i=1}^Np_i \log\biggl(
\frac{p_i}{s_i}\biggr)=D(\mathbf{p}_N \| \mathbf{s}_N) .
\]
\end{result}

 The same Bregman divergence results from the separable
component function $\phi(p) =\varphi_2(p)$, where $\varphi_2(p)\equiv p \log(p) +(1-p)$.

As to extropy, again in the separable case consider the component
function $\phi^c(p) =\varphi_1^c(p)$, where $\varphi_1^c(p)\equiv
\varphi_1(1-p)=(1-p) \log(1-p) $. This identifies the vector Bregman
function as $\Phi^c(\mathbf{p}_N) = -J(\mathbf{p}_N)$.  Since
${\phi^{c}}'(p) = -\log(1-p) - 1$, another direct application of
(\ref{eq:separableBreg}) yields a complementary result regarding $D^c(\cdot
\|\cdot)$:

\begin{result}\label{r11}
The Bregman divergence associated with
$\Phi^c(\mathbf{p}_N)=-J(\mathbf{p}_N)$ is
\begin{eqnarray*}
d_{\Phi^c}(\mathbf{p}_N, \mathbf{s}_N) &=& \sum
_{i=1}^N(1-p_i)\log\biggl(
\frac{1-p_i}{1-s_i}\biggr)
\\
&=& D^c(\mathbf{p}_N \| \mathbf{s}_N) .
\end{eqnarray*}
 \end{result}

 This same Bregman divergence also results from the Bregman
function associated with  $\phi^c(p) = \varphi_2(1-p)$, where
$\varphi_2(1-p)\equiv (1-p) \log(1-p) + p$.

 It is clear that
the duality of entropy and extropy persists through the
representation of relative (entropy, extropy) as complementary
Bregman divergences for dual Bregman functions.

\subsection{(Relative Entropy, Relative Extropy)
for Continuous Densities}\label{s5.2}

The unification of the general theory of directed distances
formulated via Bregman functions provides the representations of
entropy and extropy for continuous densities as well.  Similar to
the form of the separable Bregman divergence between two vectors,
the Bregman directed distance between two density functions $f(\cdot)$
and $g(\cdot)$ defined on $[x_1,x_N]$,
associated with a function $\phi(\cdot)$, is denoted by  $B_{\phi}(f,g)$, defined to equal
\[
\int_{x_1}^{x_N}\bigl\{\phi\bigl(f(x)\bigr)-\phi
\bigl(g(x)\bigr)-\phi'\bigl(g(x)\bigr)\bigl[f(x)-g(x)\bigr]\bigr\}
\,dx .
\]
The function $\phi\dvtx (0,\infty)\rightarrow\Re$ should be
differentiable and strictly convex, and the limits $\lim_{x
\rightarrow 0}\phi(x)$ and $\lim_{x \rightarrow 0} \phi'(x)$ must
exist (in some topology), but not necessarily be finite.  See
\citet{FrSG08}, page~1681,
and
\citet{Bass13}, page~623.
Moreover, the integral operation is constrained to be an integration
over the two functions' common domain.

 It is well known that
when $\phi(f) = \varphi_1(f)\equiv f  \log(f)$, or $\phi(f)
=\varphi_2(f)\equiv f  \log(f)+(1-f)$, specifying a convex function
defined on $[0, +\infty)$ which satisfies these conditions, then
\[
B_{\phi}(f,g) = \int_{x_1}^{x_N} f(x)\log
\biggl(\frac{f(x)}{g(x)} \biggr)\,dx .
\]
This Bregman directed distance is known as the relative entropy between
the two densities, denoted by $d(f\|g)$.

To specify the relative extropy between two densities $f(\cdot)$ and
$g(\cdot) $, we begin by recalling the relative extropy between the
mass functions $\mathbf{p}_N$ and $\mathbf{s}_N$ as represented in the
equality following (\ref{eq:DcRepresentation}):
%
\begin{eqnarray}
\label{eq:after10rep}\quad D^c(\mathbf{p}_N \| \mathbf{s}_N)
&=& J(\mathbf{s}_N) - J(\mathbf{p}_N) \nonumber\\[-8pt]\\[-8pt]
&&{}+\sum
_{i=1}^N (p_i-s_i)
\log(1-s_i) .\nonumber
\end{eqnarray}
 On the basis of its Maclaurin series expansion, the
function $(1-p_i) \log(1-p_i) \approx -p_i + \frac{1}{2} p_i^2$ when
$p_i$ is small and, thus, $J(\mathbf{p}_N) =
-\sum_{i=1}^N(1-p_i)\log(1-p_i)  \approx   1 - \frac{1}{2}
\sum_{i-1}^N p_i^2$ when $\max  p_i$ is small. Of course, a similar
result pertains to $J(\mathbf{s}_N)$.  Moreover, the common recognition
that $\log (1-s_i) \approx -s_i$ for small values of $s_i$ yields
$(p_i-s_i)\log(1-s_i) \approx -p_i s_i + s_i^2$. Applying these two
approximations (which agree with the bivariate Maclaurin series
expansion through order 3) to equation (\ref{eq:after10rep}) yields
the surprising recognition that
%
\begin{equation}
\label{eq:approxDc} D^c(\mathbf{p}_N \| \mathbf{s}_N)
\approx \frac{1}{2} \sum(p_i - s_i)^2
\end{equation}
 when  both  $\max  p_i$ and $\max  s_i$  are small.

  This is one-half the usual squared Euclidean distance between the
vectors $\mathbf{p}_N$ and $\mathbf{s}_N$; moreover, it  is also the
Bregman divergence associated with the  component function
$\phi(p)=\varphi_3(p)\equiv-p+\frac{p^2}{2}$ or
$\phi(p)=\varphi_4(p)\equiv\frac{p^2}{2}$.

A sensible definition for the relative extropy between two densities
arises from each of two consequences of this fact. First,
replacing the two component arguments of $D^c(\mathbf{p}_N \| \mathbf{s}_N)$ in (\ref{eq:approxDc}) by $p_i = f(x_i)\triangle x$ and $s_i
= g(x_i)\triangle x$, as when motivating the definitions of
differential (entropy, extropy) in Appendix \ref{appD}, we find that
\[
\lim_{\triangle x \rightarrow 0} \frac{D^c(\mathbf{p}_N\|\mathbf{s}_N)}{\triangle
x} = \frac{1}{2}\int
_{x_1}^{x_N}\bigl[f(x)-g(x)\bigr]^2\,dx .
\]
 Second, this same formulation arises from evaluating the
Bregman divergence between the densities $f(\cdot)$ and $g(\cdot)$ over a
closed interval $[x_1, x_N]$ corresponding to either of the\vspace*{1pt} convex
functions
$\phi(f)= \varphi_3(f)$  or $\phi(f)=\varphi_4(f)$, where $\varphi_3(f)=-f+\frac{1}{2}f^2$ and $\varphi_4(f) = \frac{1}{2}f^2$ ,  viz.,
\[
B_{\phi}(f,g)= \frac{1}{2}\int_{x_1}^{x_N}
\bigl[f(x)-g(x)\bigr]^2\,dx .
\]
 Motivated by these two results, we define the following:

\begin{definition}\label{d6}
The \emph{relative extropy in a
density} $f(\cdot)$ relative to $g(\cdot)$ defined over $[x_1, x_N]$ is
\[
d^c(f\|g) \equiv \frac{1}{2}\int_{x_1}^{x_N}
\bigl[f(x)-g(x)\bigr]^2\,dx .
\]
\end{definition}

 The status of relative entropy and half the $L_2$ metric as
Bregman divergences are well known. However, they have never been
recognized heretofore as formulations of the complementary duals,
entropy and extropy. For example,
\citeauthor{CeZe97} [(\citeyear{CeZe97}), page~33]
  refer to
these as ``the most popular Bregman functions,'' without any hint
how they are related.

We should expressly clarify that the
duality of entropy and extropy we are touting is distinct from the
Legendre duality between points and lines that underlies the general
structure of Bregman divergences. See
\citeauthor{BoNR10} (\citeyear{BoNR10}), Section~2.2.
Ours is a content-based duality that derives from their symmetric
co-referential relation which we exposed following Result \ref{r4} in
Section~\ref{s4}.  In this regard it is quite surprising and provocative
that half the squared $L_2$ distance (the relative extropy between
two densities) arises as the dual of the entropic norm of
Kullback--Leibler.

It is satisfying that a final result
codifies the definitions of Shannon's ``analogue'' differential
entropy function $h(f)\equiv-\int_{x_1}^{x_N} f(x) \log(f(x)) \,dx$
and our differential extropy function $j(f)\equiv - \frac{1}{2}
\int  f^2(x)\, dx  $ (discussed in Appendix \ref{appD}) as a special case of
their relative
measures with respect to a uniform density:

\begin{result}\label{r12}
Suppose $f(\cdot)$ is any density defined on
$[x_1, x_N]$ and that $u(x) = (x_N - x_1)^{-1}$ is a uniform
density.  Then the relative (entropy, extropy) pair identify the
differential (entropy, extropy) forms
\[
d(f\|u) = h(u) - h(f)
\]
and
\[
d^c(f\|u) = j(u) - j(f) .
\]
\end{result}

 Recalling from Result \ref{r7} the relation
of relative extropy $D^c(\mathbf{p}_N\|\mathbf{s}_N)$ to the relative
entropy in the complementary mass functions via $D(\mathbf{q}_N\|\mathbf{
t}_N)$, it would seem natural to search for the general relative
extropy measure between any two densities by searching for an
appropriate complementary density to a density.  As it turns out,
such a search would be chimeric because the complementary density to
every density is identical \ldots the uniform density!  This can be
recognized by examining the complementary mass function $\mathbf{q}_N
\equiv (\mathbf{1}_N - \mathbf{p}_N)/(N-1)$.  In the limiting process we
have devised, the value of $N$ increases while the maximum value of
the $\mathbf{p}_N$ vector becomes small, with each component of $\mathbf{p}_N$  converging toward zero.  In the process, each of their
complementary p.m.f.\vspace*{1pt} components becomes indistinguishable from
$\frac{1}{N}$.  Thus, the complementary density values become
uniform everywhere.

 This argument also implies that the values
of the two expectations in the limiting equation of Result \ref{r8},
$E_{\mathbf{p}_N} [\log(t^o(X))]$ and $E_{\mathbf{s}_N}[\log(t^o(X))]$, both
become indistinguishable from $\log(N)$ as $N$ increases.  This is
the entropy of a uniform p.m.f. Thus, in the limit their \emph{difference} equals $0$.

\section{Statistical Application  to Proper Scoring Rules}\label{s6}

Our discovery of extropy was stimulated by a problem that arises in
the application of the theory of proper scoring rules for
alternative forecast distributions.  These functions are the central
construct of a subjectivist statistical practice used to evaluate the
relative quality of different asserted distributions. A proper
scoring rule $S(\mathbf{p}_N, X=x^o)$ is a function of both the p.m.f.
assertion and the observation value, with the property that the
expected scoring function value (with respect to the asserted p.m.f.
$\mathbf{p}_N$) exceeds the expected score to be achieved by any other
p.m.f. The application of such rules for theory comparison is said to
promote honesty and accuracy in one's assessment of a p.m.f. to assert.
There are many proper scoring functions. \citet{DeGr84} discusses the
relation of the various
scoring functions to differing utility functions.

Proper scoring rules were the last applied statistical topic
addressed in the publications of \citet{Sava71}. They are presented
systematically and promoted in the text of \citet{Lad96}. Theory and
applications over the past half century have been reviewed by
\citet{GnRa07}. The log probability for the observed outcome of $X =
x^o$ is widely considered to be an eminent proper scoring rule and
has been used extensively: $S_{\log}(\mathbf{p}_N, X=x^o)  =
\sum_{i=1}^N (X=x_i)  \log p_i = \log(p^o)$. This score has long
been recognized to be the unique proper scoring rule for
distributions that are a function \emph{only} of the observed value of
$X=x^o$, irrespective of the probabilities assessed for the
``unobserved'' possibilities of $X$.  See \citet{ShAE66}  and
\citet{Bern79}.  The probability assessor's expected logarithmic
score equals the negentropy in the assessed distribution:
\[
E_{\mathbf{p}_N} \bigl[S_{\log}\bigl(\mathbf{p}_N,
X=x^o\bigr) \bigr]=\sum_{i=1}^N
p_i \log(p_i) .
\]

 It now appears that the logarithmic score's claim to fame
should be viewed as a weakness rather than a virtue,  for it
provides an \emph{incomplete} assessment of the probabilities
composing $\mathbf{p}_N$. The recognition of extropy as the
complementary dual of entropy plays on the fact that the observation
of $X = x^o$ is concomitant with the observations that $X \neq x_i$
for \emph{every other} $x_i$ in the range of $X$ that is different
from $x^o$. Probabilities for these \emph{observed} negated events
are inherent in the assertion of $\mathbf{p}_N$, yet the logarithmic
scoring function ignores them. The \emph{total logarithmic scoring
rule} has been proposed to address this issue:
\begin{eqnarray*}
&&S_{\mathrm{Totallog}}\bigl(\mathbf{p}_N, X=x^o\bigr)
\\
&& \quad\equiv \sum_{i=1}^N
(X=x_i) \log p_i + \sum_{i=1}^N
(X \neq x_i) \log (1-p_i) .
\end{eqnarray*}
Evidently, the expectation of this score equals the
negentropy plus the negextropy of the distribution:
\begin{eqnarray*}
&&E_{\mathbf{p}_N} \bigl[S_{\mathrm{Totallog}}\bigl(\mathbf{p}_N,
X=x^o\bigr) \bigr]
\\
&&\quad=\sum_{i=1}^N p_i
\log(p_i)+\sum_{i=1}^N
(1-p_i) \log(1-p_i) .
\end{eqnarray*}
 Moreover, each component sum and any positive linear
combination of the two components of the Total log score is a proper
score as well.

 A preliminary report by \citet{LaSA12} investigates the
importance of this issue in an application scoring alternative
forecasting distributions for daily stock prices (Agr{\`o}, Lad and Sanfilippo, \citeyear{AgLS10}).
The distributions considered differ in the attitudes they portray
toward tail area probabilities,  and the two components of the
Total log score assess the expected price and the tail area
probabilities in different ways. The international financial
collapse of recent years has accentuated an awareness of the
importance of evaluating probabilities for extreme events that
seldom occur, even when they don't occur.  One of the major insights
the report provides is that the quadratic scoring rule for
distributions should be considered not as an alternative to the
usual log score but as a complement. For while the utility of a
price forecast surely does derive from decisions that depend on the
expected prices, it also hinges on the level of insurance cover
suggested by the forecasting distribution to protect against extreme
outcomes. It should become standard practice to evaluate the
logarithmic score and the quadratic score in tandem. This conclusion
derives from the same logic we have used in this article in
identifying the squared $L_2$ distance as the extropic complement to
the Kullback--Leibler formulation of relative entropy.

Further
applications of this notion are already being promoted. An extension
of the total log proper scoring rule for probability distributions
to \emph{partial probability assessments} has been given in
\citet{CaRV10}  as a discrepancy measure between a conditional
assessment and the class of unconditional probability distributions
compatible with the assessments that \emph{are} made. Taking the work
of \citet{PSLO09} as a starting point, \citet{GiSa11a} use the
extension of a scoring rule to partial assessments while analyzing
the Total log score as a particular Bregman divergence.
\citet{BiGS12}  address the case of conditional prevision
assessments.

\section{Concluding Discussion}\label{s7}
What's in a name?  We are aware of prior uses of the word
``extropy,'' documented in both the \emph{Online Oxford English
Dictionary} and in \emph{Wikipedia}.  In one usage it seems to have
arisen as a metaphorical term rather than a technical term, naming a
proposed primal generative natural force that stimulates order
rather than disorder in both physical and informational systems.  In
the other usage within a technical context, ``extropy''  has
apparently had some parlance being used interchangeably with the
more commonly used ``negentropy,'' the negative scaling of entropy.
Neither usage of ``extropy'' appears to be very common. While we are
not stuck on this particular word, the information measure we have
introduced in this article seems aptly to merit the coinage of
``extropy.'' Whereas entropy is recognized as minus the expected log
probability of the occurring value of $X$ (a measure which could be
considered ``interior'' to the observation $X$), our proposed
extropy is derived from the expected log nonoccurrence probability
for the partition event that \emph{does} occur less the sum of log
nonoccurrence probabilities, that is, $ \sum_{i=1}^N p_i\log(1-p_i)
-\sum_{i=1}^N \log(1-p_i)$. This could be considered to be a measure
``exterior'' to the observation $X$. The exterior measure of all the
nonoccurring quantity possibilities is complementary to the entropy
measure of the unique occurring possibility.  Together, in their
joint assessment of the information inhering in a system of
probabilities, entropy and extropy identify what many people think
of as yin and yang, and what artists commonly refer to as positive
and negative space.

A word is in order about concerns of
mathematical statisticians regarding the limitations of the theory
of continuous information measures. These typically revolve upon
measurability conditions and the limitation of continuous extropy to
$L_2$ densities.  In our present digital age, the time has surely
come for statistical theorists to come to grips with the fact that
\emph{every} statistical measurement procedure in any field
whatsoever is actually limited to a finite and discrete set of
possible measurement values.  No one has ever observed a real-valued
measurement of anything.  The actual application of statistics to
inference or estimation problems involves only discrete finite
quantities.  Of course, continuous mathematics is useful for
approximate computations in situations of fine measurements.
However, such approximations need not require every imaginable
feature of mathematical structures for real computational problems.
This outlook stands in contrast to received attitudes from earlier
centuries.  These were based on the notion that reality is actually
continuous and that numerical methods of applied mathematics can
only yield discrete approximations.  We ought to recognize that such
notions are now outdated.

The statistical application to proper
scoring rules that we outlined in Section~\ref{s6} is one of many areas of
possible relevance of our dual construction.  In any commercial or
scientific arena in which entropic computations have become
standard, such as astronomical measurements of heat distribution in
galaxies, the insights provided by extropic computations would be
well worth investigating. Unrecognized heretofore, the relevance of
the duality may lie hidden in applications already conducted and
may become apparent more widely now that it is recognized. For
example, terms comprising the \emph{difference} of extropy from
entropy arise in a representation of the Bethe free energy and the
Bethe permanent in
\citeauthor{Vont12} [(\citeyear{Vont12}), pages 7--8],
 though they are not recognized there as such. Even earlier,
the Fermi--Dirac entropy function applied in nuclear physics
specifies the \emph{sum} of extropy and entropy as its Bregman
divergence without recognizing the duality of the two components.
See  \citet{FuMi12}.  Given the broad range of
applications of entropy on its own over the past half century, we
suspect that the awareness of extropy as its complementary dual will
raise as many new interesting questions as it answers.

\begin{appendix}
\section{Entropy \texorpdfstring{$\geq$}{>=} Extropy}\label{appA}

\begin{figure}

\includegraphics{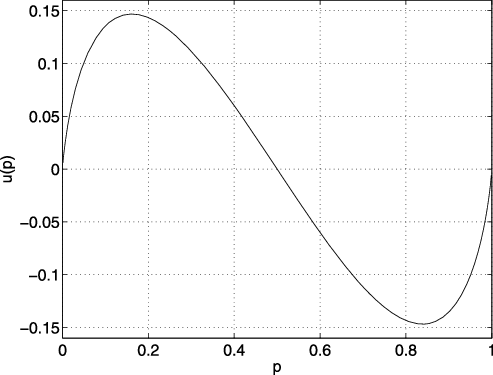}

\caption{The function $u(p)$.}
\label{fig:u}
\end{figure}

  Let $X$ be a random quantity  with a finite
discrete realm of possibilities $\{x_1,x_2,\ldots,x_N\}$
with probability masses $p_i$, with $p_i=P(X=x_i)$, $i=1,\ldots, N$.
We recall that $H(X)=-\sum_{i=1}^N{p_i}\log(p_i)$  and
$J(X)=-\sum_{i=1}^N(1-p_i)\log(1-p_i)$.
 We consider the
following real functions defined on $[0,1]$:
\begin{eqnarray*}
\varphi_1(p)&=&p \log(p) , \quad\mbox{with } 0 \log(0)\equiv 0 ;
\\
\varphi_1^c(p)&=&\varphi_1(1-p) ;
\\
u(p)&=&-\bigl(\varphi_1(p)-\varphi_1^c(p)
\bigr)
\\
&=&-p \log(p)+(1-p) \log(1-p) .
\end{eqnarray*}
The function $u(p)$ satisfies the following properties (see Figure~\ref{fig:u}):
\begin{enumerate}
\item $u(p)=0$ iff $[p=0$, or $p=1$ or $p=\frac{1}2]$;
\item $u(p)> 0$ iff $0 < p < \frac{1}2$;
\item $u(p)< 0$ iff $\frac{1}2 < p < 1$;
\item $u(1-p)=-u(p)$, for all  $ p\in [0,1]$;
\item $u(p)$ is strictly concave in   $[0,\frac{1}{2}]$,  that is, for any given pair
$(p_1,p_2)$ with $0 \leq p_1 < p_2 \in (0,\frac{1}2]$,  and for
any given $\alpha \in (0,1)$, we have
\[
u\bigl(\alpha p_1 +(1-\alpha) p_2\bigr)> \alpha
u(p_1)+(1-\alpha)u(p_2) .
\]
\end{enumerate}
By exploiting the function $u(p)$, it is evident that
\[
H(X)-J(X)=\sum_{i=1}^N
u(p_i).
\]
 This difference is permutation
invariant with respect to the components $p_i $.

We observe that for any $N>1$,  if there exist $i\in\{1,2,\ldots,
N\}$ such that $p_i=0$, then by considering
 an arbitrary quantity $Y$ with  a realm of cardinality \mbox{$N-1$} and probability masses $(p_1,p_2,\ldots,\break p_{i-1},p_{i+1}, \ldots, p_N)$
 we are ensured that
\[
H(X)=H(Y) \quad\mbox{and}\quad J(X)=J(Y).
\]
We have the following result:

Let $X$ be a finite random quantity, with realm $\{x_1,x_2,\ldots,x_N\}$ and probability masses 
$(p_1,p_2,\ldots,\break p_N)$  such that $p_i>0$, for $i=1,2,\ldots,N$, we have the following:

 (a) $H(X)=J(X)$ if $N\leq 2$;

(b) $H(X)>J(X)$ if $N > 2$.

 \textit{Case} (a). If $N=1$, we trivially  have $H(X)=J(X)=0$ and,  if $N=2$, it is $H(X)=J(X)=-p_1\log(p_1)-(1-p_1)\log(1-p_1)$.

 \textit{Case} (b).
We distinguish two alternatives:  (b1)  $p_i\leq \frac{1}2$,
$i=1,2,\ldots,N$; and (b2) $p_i> \frac{1}{2}$ for only one index~$i$.

\begin{figure*}[t]

\includegraphics{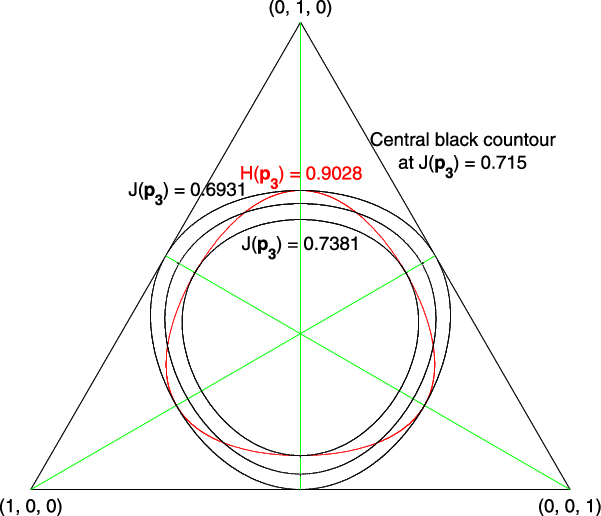}

\caption{The iso-entropy contour, $H(\mathbf{p}_3) = 0.9028$, intersects with each of the
inscribed and the exscribed iso-extropy contours at three points, and
it intersects with any intermediate iso-extropy contour at six points.
The three lines bisecting the
vertex angles partition the unit-simplex into six symmetric permutation kernels.}
\label{fig:entextpermutclass}
 \end{figure*}

 \textit{Case} (b1). By the hypotheses, for each $i$,  $0<p_i\leq
\frac{1}{2}$ and $\sum_{i=1}^N p_i = 1$.  It follows from Properties
1 and 2 of the function $u(p)$ that
\[
H(X)-J(X)=\sum_{i=1}^N
u(p_i)>0 .
\]

\textit{Case} (b2). To begin, suppose that $N=3$. Without loss of generality,
we can assume $p_3>\frac{1}2$, because of the permutation invariance
of $u(\cdot)$; consequently, $0<p_1+p_2<\frac{1}{2}$. Now from Property
4 we deduce
\[
u(p_3)=-u(1-p_3)=-u(p_1+p_2).
\]
Then  statement
\[
H(X)-J(X)=u(p_1)+u(p_2)-u(p_1+p_2)>0
\]
 amounts to
$u(p_1)+u(p_2)>u(p_1+p_2)$.\vspace*{1pt} Since $u(p)$ is strictly concave over
the interval $[0,\frac{1}{2}]$ (see Property~5)  and $u(0)=0$, we
have
%
\begin{eqnarray}
\label{eq:up1} \hspace*{40pt}u(p_1)&=&u \biggl(\frac{p_2}{p_1+p_2}0 +
\frac{p_1}{p_1+p_2}(p_1+p_2) \biggr)
\nonumber
\\&>&\frac{p_2}{p_1+p_2}u(0)+\frac{p_1}{p_1+p_2}u(p_1+p_2) \hspace*{-40pt}
\\
&=&
\frac{p_1}{p_1+p_2}u(p_1+p_2)
\nonumber
\end{eqnarray}
and
%
\begin{eqnarray}
\label{eq:up2}  \hspace*{21pt}u(p_2)&>&\frac{p_1}{p_1+p_2}u(0)+\frac{p_2}{p_1+p_2}u(p_1+p_2)
 \hspace*{-21pt}\nonumber
\\[-8pt]
\\[-8pt]
&=& \frac{p_2}{p_1+p_2}u(p_1+p_2).
\nonumber
\end{eqnarray}

From (\ref{eq:up1}) and (\ref{eq:up2}) it follows
$u(p_1)+u(p_2)>u(p_1+p_2)$ and then  $H(X)-J(X)>0$.

Generally, let $N>2$. Again\vspace*{1pt} without loss of generality, we can assume
$p_N>\frac{1}2$. We have
\[
u(p_N)=-u(1-p_N)=-u(p_1+p_2+
\cdots+p_{N-1}).
\]
For each $i=1,\ldots, N-1$, it is easy to see that
%
\begin{eqnarray}
u(p_i)&>&\frac{p_i}{p_1+p_2+\cdots+p_{N-1}}\nonumber\\[-8pt]\\[-8pt]
&&{}\cdot u(p_1+p_2+
\cdots+p_{N-1}),\nonumber
\end{eqnarray}
 because of the concavity of $u(\cdot)$.

 Finally, we have
\begin{eqnarray*}
 &&H(X)-J(X) \\
 &&\quad= \sum_{i=1}^N
u(p_i)
\\
&&\quad = \sum_{i=1}^{N-1}
u(p_i)-u(p_1+p_2+\cdots+p_{N-1})\\
&&\quad>0
.
\end{eqnarray*}

\section{The Range of Extropy Values That Share an Entropy}\label{appB}

In the same observational context as Figure~\ref{fig:entandextcontours},
Figure~\ref{fig:entextpermutclass}
displays a single entropy contour at the value $H(\mathbf{p}_3) =
0.9028$.  Inscribed and exscribed are the maximum and minimum extropy
contours that intersect with it. Each of these extreme extropy
contours has three intersection points with the entropy contour, and
the p.m.f. that each of these points represents has two equal
components. So the three triples constituting the mass function
intersection points on both the max and the min $J$ contours are
permutations of one another.  The intermediate extropy contour
intersects the $H(\mathbf{p}_3) = 0.9028$ contour at six points, the six
permutations of a $\mathbf{p}_3$ vector with \emph{three distinct}
components. Both the $H(\cdot)$ and $J(\cdot)$ functions are permutation
invariant. In higher dimensions, the intersection of $H(\mathbf{p}_N)$
and $J(\mathbf{p}_N)$ contours yields surfaces in $(N-2)$ dimensions
that are symmetric across the permutation kernels of the
unit-simplex $\mathbf{S}^{N-1}$.

When the entropy is calculated for any assemblage such as the heat
distribution for a galaxy of stars, a companion calculation of the
extropy would allow us to complete our understanding of the
variation inherent in its empirical distribution.  The extropy value
completes the measure of disorder in the array, placing it within
the extremes that are possible for the calculated entropy value.

\section{Isocontours of \texorpdfstring{$H(\cdot)$}{$H(cdot)$} Generate Isocontours
of \texorpdfstring{$J(\cdot)$}{$J(cdot)$} via Result \texorpdfstring{\protect\ref{R5}}{5}}
\label{appC}

As a numerical and geometrical example, consider again Figure~\ref{fig:entandextcontours} in the context of the following
computational results.  These need to be compared with the points
they represent in the figure as you go. To begin, notice that
$H(\frac{1}4,\frac{1}2,\frac{1}4) = 1.0397$ and\vspace*{1pt}
$J(\frac{1}4,\frac{1}2,\frac{1}4) = 0.7781$ identify the points at the apex
of specific isoentropy and isoextropy contours from your perspective
as you view the left and right sides of Figure~\ref{fig:entandextcontours}. Both of these
contours lie precisely on and are tangent to the triangular
sub-simplex $\mathbf{S}_c$ that is inscribed within the unit-simplex
$\mathbf{S}^2$ in Figure~\ref{fig:entandextcontours}(left)
and Figure~\ref{fig:entandextcontours}(right). Result \ref{R5} tells us that the
source of this isoextropy contour on the right is the higher level
isoentropy contour $H = 1.082$ that contains the point $\mathbf{q}_3 =
(\frac{3}8,\frac{1}4,\frac{3}8)$ at the bottom of this entropy contour.
This is the mass function complementary to $\mathbf{p}_3 =
(\frac{1}4,\frac{1}2,\frac{1}4)$. Computationally, $J(\mathbf{p}_3 =
(\frac{1}4,\frac{1}2,\frac{1}4))  =  0.7781  = { 2 [ H(\mathbf{q}_3 =
(\frac{3}8,\frac{1}4,\frac{3}8))  -   \log(2) ] }   =   2  [  1.0822
- 0.6931 ]$,\vspace*{1pt} as prescribed by Result \ref{R5}.  Transformed into an
isoextropy contour, this isoentropy contour containing $H(\mathbf{q}_3)
= 1.0822$ is flipped and expanded to represent $J(\mathbf{p}_3) =
0.7781$. If we would \emph{begin} with a consideration of the \emph{entropy} contour containing $H(\frac{3}8,\frac{1}4,\frac{3}8) = 1.0822$,
regarding \emph{this} triple as $\mathbf{p}_3$, we would find its dual
extropy contour is denominated $J = 0.8033$, containing the member
$J(\frac{3}8,\frac{1}4,\frac{3}8) = 0.8033$.\vspace*{1pt} These two contours are
precisely inscribed in the sub-sub-simplex $\mathbf{S}_{cc}$ which is
inlaid within $\mathbf{S}_c$ in Figure~\ref{fig:entandextcontours}(left)
and Figure~\ref{fig:entandextcontours}(right). This
visualization completes our understanding of extropy as the
complementary dual of entropy.\looseness=1

\section{Differential Entropy and Extropy for Continuous Densities}
\label{appD}

We begin this exposition by reviewing how the analogical character
of Shannon's differential entropy measure for a continuous density
derives from its status as the limit of a linear transformation of
the discrete entropy measure.

\subsection{Shannon's Differential Entropy: \texorpdfstring{$-\int f(x)\log f(x)\,dx$}{$-int f(x)\log f(x)\,dx$}}

For the following simple exposition of Shannon's considerations,
presume again that the range of a quantity $X$ is $\{x_1, \ldots,
x_N\}$ and that the values of $x_1$ and $x_N$ are fixed.  For each
larger value of $N$, presume that more elements are included
uniformly in the interval between them and that the $p_i$ values
are refined in such a way that the maximum $p_i$ value reduces
toward $0$. Now define $\triangle x   \equiv(x_N - x_1)/(N-1)$ for
any specific $N$, and define $f(x_i) \equiv p_i/\triangle x$. In
these terms, the entropy $H(\mathbf{p}_N)$ can be expressed as
%
\begin{eqnarray}
\label{eq:H(p)representation}  \hspace*{26pt}H(\mathbf{p}_N) &=& -\sum p_i
\log(p_i) \hspace*{-26pt}
\nonumber
\\
&= & -\sum f(x_i) \triangle x \log\bigl(f(x_i)
\triangle x\bigr)
\\
&= &-\sum f(x_i) \log\bigl(f(x_i)\bigr)
\triangle x - \log(\triangle x) .
\nonumber
\end{eqnarray}
 Thus, the entropy measure $H(\mathbf{p}_N)$ is unbounded as $N$
increases, with $\triangle x \rightarrow 0$.  However, the
summand $-\sum f(x_i) \log(f(x_i)) \triangle x$ on  (\ref{eq:H(p)representation})  is merely a location transform of the
entropy $-\sum p_i \log(p_i)$, shifting only by $\log(\triangle x)$
which is finite for any $N$. The limit of the relocated entropy
expression suggests Shannon's definition of the continuous analogue:

\begin{Definition}\label{dD.1}
 The \emph{differential entropy} of a
density $f(\cdot)$ over the interval $[x_1, x_N]$ is defined as
\begin{eqnarray*}
h(f)
&\equiv& -\int_{x_1}^{x_N} f(x) \log  \bigl(f(x)  \bigr)\, dx \\ &=&
\lim_{\triangle x \rightarrow 0} \bigl[H(\mathbf{p}_N) +\log  \triangle x\bigr].
\end{eqnarray*}
\end{Definition}

Shannon himself noted that this analogous measure loses the absolute
meaning that the finite measure enjoys, because its value must be
considered relative to an assumed standard of the coordinate system
in which the value of the variable is expressed.  If the variable
$X$ were transformed into $Y$, then the continuous measure of the
differential entropy $h_Y(f(\cdot))$ needs to be adjusted from
$h_X(f(\cdot))$ by the Jacobian of the specific transformation. He
suggested, however, that the continuous analogue retains its value as
a \emph{comparative} measure of the uncertainties contained in two
densities because they would both be affected by the transformation
in the same way. In any case, the characterization of \emph{relative}
entropy, which we address in Section~\ref{s5.1}, has been found to
circumvent the invariance problem. See the discussion in
\citet{Cati12}, page~85.  We shall now examine differential extropy
in the style suggested by Shannon's argument.

\subsection{Motivating the Differential Extropy Measure as \texorpdfstring{$-\frac{1}{2}\int f^2(x)\,dx$}{$-\frac{1}{2}int f^2(x)\,dx$}}

At first sight, the extropy measure $-\sum (1-p_i) \*\log(1-p_i)$
appears problematic:  if each $p_i$ were simply replaced by a
density value $f(x)$, the measure would not be defined when $f(x)
> 1$, which it may.  However, the situation clarifies by expanding
$(1-p_i) \log(1-p_i)$ through three terms of its Maclaurin series
with remainder:    $(1-p_i) \log(1-p_i) =  - p_i   + \frac{p_i^2}{2} + \frac{p_i^3}{6(1-r_i)^2}$ for some $r_i  \in (0,p_i)$. Summing these expansion
terms over $i = 1,\ldots,N$ shows that when the range of possibilities
for $X$ increases (as a result of larger $N$) in such a way that
$\triangle x \rightarrow 0$ and $ \max_{i=1}^{N}  p_i$
decreases toward $0$, the extropy measure becomes closely
approximated by $ 1  -  \frac{1}{2} \sum_{i=1}^N p_i^2 $.

Following the same tack as for entropy in representing $p_i$ by
$f(x_i)\triangle x$ suggests that for large $N$ the extropy measure
can be approximated by
\begin{eqnarray*}
J(\mathbf{p}_N) &\approx & 1 - \frac{1}{2} \sum
_{i=1}^N p_i^2 \quad (
\mbox{when } \max p_i \mbox{ is small})
\nonumber
\\
&= & 1 - \frac{1}{2} \sum f^2(x_i) (
\triangle x)^2
\nonumber
\\
&= & 1 - \frac{\triangle x}{2} \sum f^2(x_i)
\triangle x .
\end{eqnarray*}
 This approximation is merely a location and scale
transformation of  $-\frac{1}{2} \sum f^2(x_i) \triangle x
$.  In the same spirit as for differential entropy, the measure of
differential extropy for a continuous density can well be defined
via the limit of $J(\mathbf{p}_N)$ as $N$ increases
in the same context as Definition~\ref{dD.1}:

\begin{Definition}\label{dD.2}
The \emph{differential extropy} of
the density $f(\cdot)$ is defined as
\[
j(f)   \equiv   -
\frac{1}{2} \int  f^2(x)
\,dx  =  \lim_{\triangle x \rightarrow 0} \bigl\{\bigl[J(\mathbf{p}_N) - 1\bigr]
/   \triangle x\bigr\}.
\]
\end{Definition}

The sum of the squares of probability masses (as well as the
integral of the square of a density) has received attention for more
than a century for a variety of reasons, but never in a direct
relation to the entropy of a distribution. Rather, it has merely
been considered to be an \emph{alternative} measure of uncertainty.
\citet{Good79}  referred to this measure as the ``repeat rate'' of a
distribution, developing an original idea of Turing.
\citeauthor{Gini12} (\citeyear{Gini12,Gini39})  had earlier proposed this measure as an
``index of heterogeneity'' of a discrete distribution, via $1-
\sum_{i=1}^N p_i^2$, deriving from the sum of the individual event
variances, $p_i(1-p_i)$. We now find that in a discrete context, a
rescaling of Gini's index is an approximation to the extropy of a
distribution when the \mbox{maximum} probability mass is small.  In a
continuous context, half the negative expected value of a density
function value is the continuous differential analogue of the
extropy measure
of a distribution that we are proposing.
\end{appendix}

\section*{Acknowledgments}
This research supported in part by Grant FFR 2012-ATE-0585 from the University of
Palermo. Thanks to Patrizio Frederic, Gianfranco
Lovison, Marcello Chiodi, Jim Dickey (RIP), James O'Malley, Mary Lou
Zeeman, Bernhard Geiger and Michele Tumminello for helpful
discussions over some years. Some of these results were
discussed as part of an invited lecture to the Brazil Bayesian
Statistics Conference, EBEB 2012 in Amparo, Sao Paulo, and in a
lecture at the Institute for Reliability and Risk Analysis of George
Washington University in October, 2012. Thanks to Marcio Diniz,
Ariel Caticha, Nozer Singpurwalla, and to two editors (current and immediate past),
an associate editor and two reviewers for helpful comments.



\end{document}